# Theoretical Limits on Time Delay Estimation for Ultra-Wideband Cognitive Radios

*Invited Paper*

Sinan Gezici[*], Hasari Celebi[♯], Huseyin Arslan[♯], and H. Vincent Poor[†]

[*] Department of Electrical and Electronics Engineering, Bilkent University, Bilkent, Ankara 06800, Turkey
[♯] Department of Electrical Engineering, University of South Florida, 4202 E Fowler Ave, Tampa, FL 33620, USA
[†] Department of Electrical Engineering, Princeton University, Princeton, NJ 08544, USA
gezici@ee.bilkent.edu.tr, {hcelebi@mail,arslan@eng}.usf.edu, poor@princeton.edu

*Abstract*— In this paper, theoretical limits on time delay estimation are studied for ultra-wideband (UWB) cognitive radio systems. For a generic UWB spectrum with dispersed bands, the Cramer-Rao lower bound (CRLB) is derived for unknown channel coefficients and carrier-frequency offsets (CFOs). Then, the effects of unknown channel coefficients and CFOs are investigated for linearly and non-linearly modulated training signals by obtaining specific CRLB expressions. It is shown that for linear modulations with a constant envelope, the effects of the unknown parameters can be mitigated. Finally, numerical results, which support the theoretical analysis, are presented.

## I. INTRODUCTION

Cognitive radio systems provide intelligent wireless communications by means of adaptation, feedback, learning, and spectrum sensing. The main characteristics of cognitive radio are related to spectrum utilization in a dispersed manner; i.e., a cognitive radio system can utilize a number of adjacent or non-adjacent frequency bands at the same time [1], [2]. In an ultra-wideband (UWB) cognitive radio system, the available spectrum has ultra-wide bandwidth [3], and the spectrum can be used in a dispersed manner [4], [5].

Cognitive radio systems with *location awareness* provide location information to the users, which facilitate location-based services and applications, and also help network optimization [6], [7]. The part of the cognitive radio system that provides location sensing is called a *cognitive positioning system* (CPS) [8]. CPSs can use Cramer-Rao lower bound (CRLB) information at the transmitter side to optimize the transmission parameters for achieving specific accuracy requirements. In [8], the CRLB for time delay estimation in multipath channels is derived, where the path delays and coefficients are assumed to be unknown. An extension of [8] is provided in [9], which obtains the frequency-domain CRLB in dynamic spectrum access systems for unknown path delays, distance-dependent coefficients, phases and frequency-dependent coefficients. Both [8] and [9] consider the available bandwidth in the form of a single piece (i.e., the whole bandwidth). However, the available bandwidths in UWB cognitive radio systems are commonly dispersed [10] and each user can transmit and receive over multiple dispersed bands. This study considers UWB cognitive radio systems that can employ various dispersed bands [11].

In this paper, CRLBs for time delay estimation are studied for UWB cognitive radio systems with dispersed frequency bands, and the effects of unknown channel coefficients and carrier frequency offsets (CFOs) are quantified. After obtaining a generic expression, specific CRLB formulas are obtained for various modulation schemes. Then, it is shown that the same lower bound can be achieved for the cases of known and unknown CFOs for linear modulation schemes. In addition, it is proven that the effects of unknown channel coefficients can be mitigated for linear modulations with a constant envelope. Finally, numerical results are presented and concluding remarks are made.

## II. SIGNAL MODEL

A UWB cognitive radio system that occupies $K$ different frequency bands, which can be adjacent or non-adjacent, is considered. The transmitter sends a signal occupying all the $K$ bands, and the receiver tries to estimate the time delay of the incoming signal. Such systems can be implemented as orthogonal frequency division multiplexing (OFDM) systems, where the sub-carriers corresponding to the unused bands are considered to have zero coefficients [12], [13]. The main disadvantage of this approach is that when the available spectrum is very dispersed, processing of very large bandwidths is required, which can increase the receiver complexity [11], [14]. Another approach to implement cognitive radio systems with dispersed spectrum utilization is to process the signal in multiple branches, as shown in Fig. 1 [11]. In this case, each branch processes one available band, which is significantly narrower than the total bandwidth. Unlike the first scenario, the system uses multiple down-conversion units in Fig. 1; hence, different carrier frequency offsets (CFOs) can be observed at the signals at different branches. Considering the receiver structure in Fig. 1, the aim is to determine the accuracy of time delay estimation in the presence of unknown channel effects and CFOs.

Since each of the $K$ frequency bands will have narrow bandwidths compared to the total system bandwidth, it is assumed that the baseband representation of the received signal in the $i$th branch can be expressed as

$$r_i(t) = \alpha_i e^{-j\omega_i t} s_i(t-\tau) + n_i(t) , \quad (1)$$

[0]This work was supported in part by the European Commission in the framework of the FP7 Network of Excellence in Wireless COMmunications NEWCOM++ (contract no. 216715) and WiMAGIC (contract no. 215167), and in part by the U. S. National Science Foundation under Grants ANI-03-38807 and CNS-06-25637.

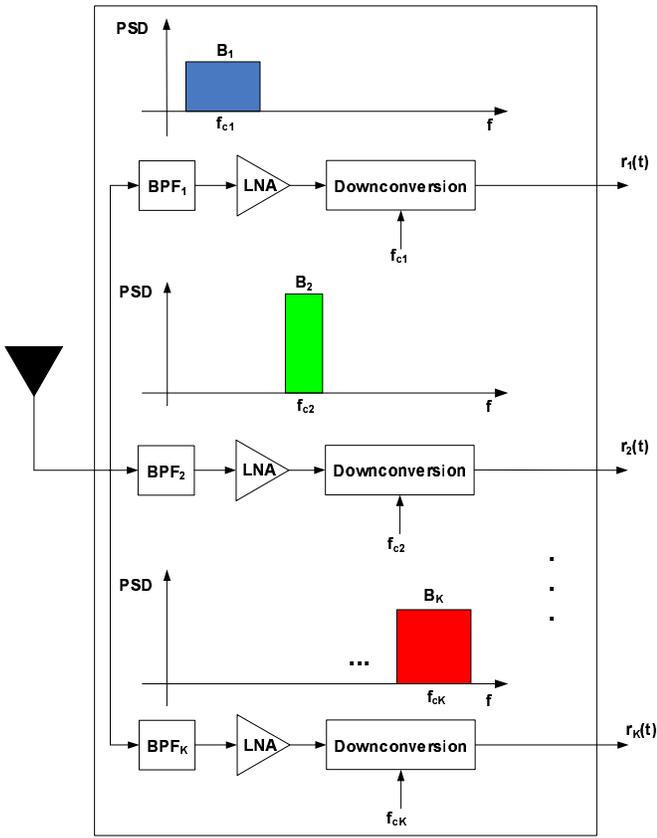

Fig. 1
BLOCK DIAGRAM OF A COGNITIVE RADIO RECEIVER WITH MULTIPLE BRANCHES.

for $i = 1, \ldots, K$, where $\alpha_i = a_i e^{j\phi_i}$ and $\omega_i$ represent, respectively, the channel coefficient and the CFO for the signal in the $i$th branch, $s_i(t)$ is the baseband representation of the transmitted signal corresponding to the $i$th band, $\tau$ is the time delay, and $n_i(t)$ is complex Gaussian noise with independent and white components, each having spectral density $\sigma_i^2$.

## III. GENERIC LIMITS

In the presence of unknown channel and CFOs, the vector of unknown signal parameters can be expressed as $\boldsymbol{\theta} = [\tau \; a_1 \cdots a_K \; \phi_1 \cdots \phi_K \; \omega_1 \cdots \omega_K]$. Then, for an observation interval of $[0, T]$, the log-likelihood function for $\boldsymbol{\theta}$ can be obtained from (1) as

$$\Lambda(\boldsymbol{\theta}) = c - \sum_{i=1}^{K} \frac{1}{2\sigma_i^2} \int_0^T \left| r_i(t) - \alpha_i e^{j\omega_i t} s_i(t-\tau) \right|^2 dt , \quad (2)$$

where $c$ is a constant that is independent of $\boldsymbol{\theta}$.

In order to calculate the CRLB on time delay estimation, the Fisher information matrix (FIM) should be calculated first [15]. From (2), it can be obtained as [11]

$$\mathbf{I} = \begin{bmatrix} \mathrm{I}_{\tau\tau} & \mathbf{I}_{\tau a} & \mathbf{I}_{\tau\phi} & \mathbf{I}_{\tau\omega} \\ \mathbf{I}_{\tau a}^T & \mathbf{I}_{aa} & 0 & 0 \\ \mathbf{I}_{\tau\phi}^T & 0 & \mathbf{I}_{\phi\phi} & \mathbf{I}_{\phi\omega} \\ \mathbf{I}_{\tau\omega}^T & 0 & \mathbf{I}_{\phi\omega}^T & \mathbf{I}_{\omega\omega} \end{bmatrix}, \quad (3)$$

with

$$\mathrm{I}_{\tau\tau} = \sum_{i=1}^{K} \gamma_i \tilde{E}_i , \quad (4)$$

$$\mathbf{I}_{aa} = \mathrm{diag}\left\{ \frac{E_1}{\sigma_1^2}, \ldots, \frac{E_K}{\sigma_N^2} \right\} , \quad (5)$$

$$\mathbf{I}_{\phi\phi} = \mathrm{diag}\left\{ E_1 \gamma_1, \ldots, E_K \gamma_K \right\} , \quad (6)$$

$$\mathbf{I}_{\omega\omega} = \mathrm{diag}\left\{ F_1 \gamma_1, \ldots, F_K \gamma_K \right\} , \quad (7)$$

$$\mathbf{I}_{\tau a} = -\left[ \hat{E}_1^{\mathrm{R}} \frac{|\alpha_1|}{\sigma_1^2} \cdots \hat{E}_K^{\mathrm{R}} \frac{|\alpha_K|}{\sigma_K^2} \right] , \quad (8)$$

$$\mathbf{I}_{\tau\phi} = -\left[ \hat{E}_1^{\mathrm{I}} \gamma_1 \cdots \hat{E}_K^{\mathrm{I}} \gamma_K \right] , \quad (9)$$

$$\mathbf{I}_{\tau\omega} = -\left[ G_1 \gamma_1 \cdots G_K \gamma_K \right] , \quad (10)$$

$$\mathbf{I}_{\phi\omega} = \mathrm{diag}\left\{ \hat{F}_1 \gamma_1, \ldots, \hat{F}_K \gamma_K \right\} , \quad (11)$$

where $\gamma_i = |\alpha_i|^2/\sigma_i^2$, $\mathrm{diag}\{x_1, \ldots, x_K\}$ represents an $K \times K$ diagonal matrix with its $i$th diagonal being equal to $x_i$, $E_i = \int_0^T |s_i(t-\tau)|^2 dt$ is the signal energy[1], $\tilde{E}_i$ is the energy of the first derivative of $s_i(t)$; i.e., $\tilde{E}_i = \int_0^T |s_i'(t-\tau)|^2 dt$, and $\hat{E}_i^{\mathrm{R}}$, $\hat{E}_i^{\mathrm{I}}$, $F_i$, $\hat{F}_i$ and $G_i$ are given, respectively, by[2]

$$\hat{E}_i^{\mathrm{R}} = \int_0^T \mathcal{R}\left\{ s_i'(t-\tau) s_i^*(t-\tau) \right\} dt , \quad (12)$$

$$\hat{E}_i^{\mathrm{I}} = \int_0^T \mathcal{I}\left\{ s_i'(t-\tau) s_i^*(t-\tau) \right\} dt , \quad (13)$$

$$F_i = \int_0^T t^2 \left| s_i(t-\tau) \right|^2 dt , \quad (14)$$

$$\hat{F}_i = \int_0^T t \left| s_i(t-\tau) \right|^2 dt , \quad (15)$$

$$G_i = \int_0^T t \mathcal{I}\left\{ s_i^*(t-\tau) s_i'(t-\tau) \right\} dt . \quad (16)$$

The CRLB for unbiased time delay estimators can be obtained from the element in the first row and first column of the inverse of the FIM in (3), i.e., $\left[\mathbf{I}^{-1}\right]_{11}$. Based on the formulas for block matrix inversion, it can be calculated as [11]

$$\mathrm{CRLB}_1 = \frac{1}{\sum_{i=1}^{K} \gamma_i \left( \tilde{E}_i - (\hat{E}_i^{\mathrm{R}})^2 / E_i \right) - \xi} , \quad (17)$$

where

$$\xi = \sum_{i=1}^{K} \gamma_i \frac{(\hat{E}_i^{\mathrm{I}})^2 F_i + E_i G_i^2 - 2\hat{E}_i^{\mathrm{I}} G_i \hat{F}_i}{E_i F_i - \hat{F}_i^2} . \quad (18)$$

In order to investigate the effects of unknown CFOs, the CRLB for time delay estimation can be calculated in the presence of known CFOs. In that case, similar derivations can be performed for the unknown parameter vector $\tilde{\theta} = [\tau \; a_1 \cdots a_K \; \phi_1 \cdots \phi_K]$, and the CRLB can be obtained as

$$\text{CRLB}_2 = \frac{1}{\sum_{i=1}^{K} \gamma_i \left( \tilde{E}_i - \hat{E}_i^2 / E_i \right)} , \quad (19)$$

where

$$\hat{E}_i = \left| \int_{-\infty}^{\infty} s_i'(t-\tau) s_i^*(t-\tau) \mathrm{d}t \right| = \sqrt{(\hat{E}_i^{\text{R}})^2 + (\hat{E}_i^{\text{I}})^2} . \quad (20)$$

---
[1]Although $E_i$ is a function of $\tau$ in general, it is not shown explicitly for convenience. The same convention is used in (12)-(16) as well.

[2]For a complex number $z$, $\mathcal{R}\{z\}$ and $\mathcal{I}\{z\}$ represent its real and imaginary parts, respectively.

Note that $\text{CRLB}_2$ in (19) is smaller than or equal to $\text{CRLB}_1$ in (17) in general, as more unknown parameters exist in the latter case.

In the extreme case in which both the channel coefficients and the CFOs are known, the unknown parameter vector reduces to $\tau$, and the CRLB expression reduces to

$$\text{CRLB}_3 = \frac{1}{\sum_{i=1}^{K} \gamma_i \tilde{E}_i} . \quad (21)$$

## IV. SPECIAL CASES

In this section, the baseband signal $s_i(t)$ in (1) is modeled to consist of a sequence of modulated pulses; i.e.,

$$s_i(t) = \sum_l d_{i,l}(t) p_i(t - lT_i) , \quad (22)$$

for $i = 1, \ldots, K$, where $d_{i,l}(t)$ denotes the complex data[3] for the $l$th symbol of signal $i$, and $p_i(t)$ represents a pulse with duration $T_i$, which is non-zero only for $t \in [0, T_i]$. In addition, it is assumed for the simplicity of the expressions that the observation interval $T$ can be expressed as $T = N_i T_i$ for an integer $N_i$ for $i = 1, \ldots, K$.

First, consider a scenario in which the signal $s_i(t)$ is a linearly modulated signal expressed as $s_i(t) = \sum_l d_{i,l} p_i(t - lT_i)$. In this case, it can be shown from (22) that $s_i'(t) s_i^*(t)$ is a real quantity; hence, $\text{CRLB}_1$ in (17) and $\text{CRLB}_2$ in (19) become [11]

$$\text{CRLB}_1 = \text{CRLB}_2 = \frac{1}{\sum_{i=1}^{K} \gamma_i \left( \tilde{E}_i - (\hat{E}_i^{\text{R}})^2 / E_i \right)} . \quad (23)$$

The result in (23) implies that for common modulation types such as pulse amplitude modulation (PAM), phase shift keying (PSK) and quadrature amplitude modulation (QAM) [16], the CRLB of time delay estimation for the case of unknown CFOs is the same as that for the case of known CFOs.

Next, consider linearly modulated pulses with a constant envelope; i.e., $s_i(t) = \sum_l d_{i,l} p_i(t - lT_i)$, with $|d_{i,l}| = |d_i| \; \forall l$. In addition, if $p_i(t)$ satisfies $p_i(0) = p_i(T_i)$ for $i = 1, \ldots, K$, the CRLBs in (17) and (19) can be calculated as

$$\text{CRLB}_1 = \text{CRLB}_2 = \text{CRLB}_3 = \frac{1}{4\pi^2 \sum_{i=1}^{K} \text{SNR}_i \beta_i^2} , \quad (24)$$

where $\text{SNR}_i = N_i |d_i|^2 \frac{|\alpha_i|^2 E_{p_i}}{\sigma_i^2}$ with $E_{p_i} = \int_{-\infty}^{\infty} p_i^2(t) \mathrm{d}t$, and $\beta_i$ is the effective bandwidth of $p_i(t)$ [11].

[3]Since data-aided time delay estimation is considered, known data symbols are assumed.

From (24), it is observed that for linear modulations with a constant envelope, such as PSK, the CRLB on time delay estimation is the same for known or unknown CFOs and/or channel coefficients when the pulses satisfy $p(0) = p(T)$, which is commonly the case in practice. Therefore, in such cases, it is possible to mitigate the effects of the unknown parameters on time delay estimation. More specifically, maximum likelihood (ML) estimators of time delay can asymptotically achieve the same CRLB for unknown CFOs and channel coefficients case as the ones in presence of CFO and/or channel coefficient information [15].

## V. NUMERICAL RESULTS AND CONCLUSIONS

In this section, numerical studies are performed in order to evaluate the CRLB expressions in the previous sections. For the cognitive system, it is assumed that all the $K$ bands in the system have the same bandwidth and the same pulse is employed for all of them; i.e., $p_i(t) = p(t)$ for $i = 1, \ldots, K$ (c.f. (22)). For the pulse shape, the following Gaussian doublet is employed

$$p(t) = A \left( 1 - \frac{4\pi t^2}{\zeta^2} \right) e^{-2\pi t^2 / \zeta^2} , \quad (25)$$

where $A$ and $\zeta$ are parameters that are used to adjust the pulse energy and the pulse width, respectively. In the following, $\zeta = 8$ ns is employed, for which the pulse width becomes approximately 20 ns, and $A$ is selected to normalize the pulse energy. In addition, the same noise spectral density is assumed for all the $K$ branches; i.e., $\sigma_i = \sigma$ for $i = 1, \ldots, K$. Finally, the system SNR is defined as the sum of the SNRs in the various branches.

In Fig. 2, the CRLB expressions in (17), (19) and (21), which are labeled as $\text{CRLB}_1$, $\text{CRLB}_2$ and $\text{CRLB}_3$, respectively, are plotted versus SNR for three different modulation types, namely 16FSK[4], 16QAM, and 16PSK. In all cases, the same modulation sequence is employed at different branches; i.e., $d_{i,l}(t) = d_l(t)$ for $i = 1, \ldots, K$, and that the channel amplitudes are normalized to unity; i.e., $|\alpha_i| = 1$ for $i = 1, \ldots, K$. In addition, the modulation sequence for each modulation type is scaled appropriately so as to equate the energy of the sequences in different scenarios. Also, there are $K = 10$ branches in the system, and $N = 2$ symbols are received at each branch. It is observed that for all the modulations, $\text{CRLB}_3 \leq \text{CRLB}_2 \leq \text{CRLB}_1$ is satisfied, since $\text{CRLB}_1$ corresponds to the case of unknown delay, channel coefficients and CFOs, $\text{CRLB}_2$ corresponds to the case of unknown delay and channel coefficients, and $\text{CRLB}_3$ corresponds to the case of unknown delay only. In other words, for the cases with fewer unknown parameters, lower CRLBs are observed. For the 16FSK modulation, all three bounds are distinct. Note that as 16FSK is a non-linear modulation, the result in (23) does not apply in this case. For the 16QAM case, $\text{CRLB}_1 = \text{CRLB}_2$ as expected from (23). However, $\text{CRLB}_3$ is lower than both, which is can happen since 16QAM is not a constant envelope modulation (c.f. (24)). Finally, for the 16PSK case, all the three bounds are equal, which verifies (24).

In Fig. 3, $N = 16$ symbols are used and the rest of the system parameters are kept the same as in the previous case.

[4]The amount of frequency shift is selected as 23.75 MHz for the FSK modulation.

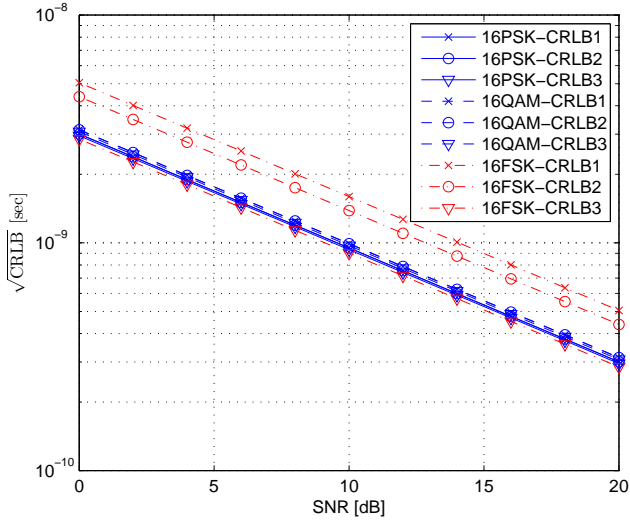

Fig. 2
$\sqrt{\text{CRLB}}$ versus SNR for $K = 10$ and $N = 2$.

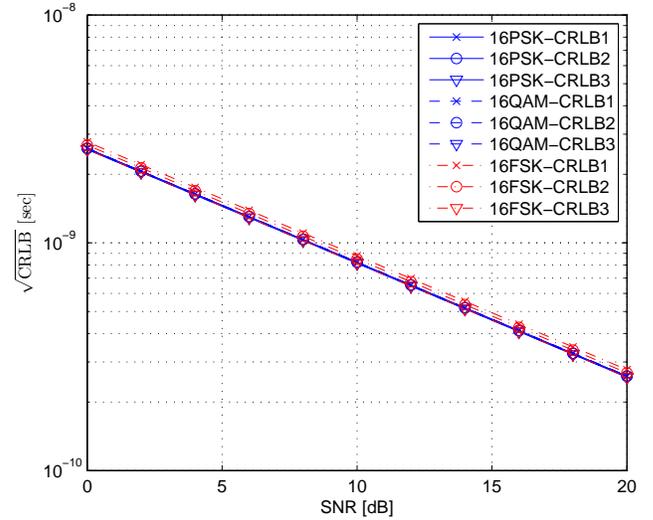

Fig. 3
$\sqrt{\text{CRLB}}$ versus SNR for $K = 10$ and $N = 16$.

It is observed that the results are similar to those in Fig. 2, except that the gap between the CRLBs is decreased, and lower CRLBs are achieved when more symbols are employed. In addition, the reduction of the gaps between $\text{CRLB}_1$, $\text{CRLB}_2$ and $\text{CRLB}_3$ implies that using a larger number of observation symbols can mitigate the effects of unknown CFOs and/or channel coefficients.

Finally, the CRLBs for different numbers of branches, which represent the numbers of available dispersed bands in the spectrum, are investigated for various SNRs. The parameters are the same as in the previous scenario, except that only 16PSK is considered here for simplicity, and the CRLB is evaluated for $K = 10, 15, 20, 25$ branches and for SNR $= 5, 10, 15$ dB. It is observed from Fig. 4 that the CRLBs decrease as the SNR increases but they are invariant to the changes in the number of branches. This is because the SNR is defined as the sum of the SNRs in the different branches; hence, the SNR per branch is reduced as more branches are employed.

## REFERENCES

[1] J. Mitola and G. Q. Maguire, "Cognitive radio: Making software radios more personal," *IEEE Personal Commun. Mag.*, vol. 6, no. 4, pp. 13–18, Aug. 1999.
[2] S. Haykin, "Cognitive radio: Brain-empowered wireless communications," *IEEE J. Select Areas Commun.*, vol. 23, no. 2, pp. 201–220, Feb. 2005.
[3] S. Gezici, Z. Tian, G. B. Giannakis, H. Kobayashi, A. F. Molisch, H. V. Poor, and Z. Sahinoglu, "Localization via UWB Radios," *IEEE Signal Processing Mag.*, vol. 22, no. 4, pp. 70–84, July 2005.
[4] J. Chiang and J. Lansford, "Use of cognitive radio techniques for OFDM ultrawideband coexistence with WiMAX," in *2005 Texas Wireless Symposium*, Austin, TX, Oct. 26-28 2005.
[5] M. E. Sahin, S. Ahmed, and H. Arslan, "The roles of ultra wideband in cognitive networks," in *IEEE International Conference on Ultra-Wideband*, Singapore, Sept. 24-26 2007, pp. 247–252.
[6] H. Celebi and H. Arslan, "Utilization of location information in cognitive wireless networks," *IEEE Wireless Commun. Mag.-Special issue on Cognitive Wireless Networks*, vol. 14, no. 4, pp. 6–13, Aug. 2007.
[7] ——, "Enabling location and environment awareness in cognitive radios," *Elsevier Computer Communications (Special Issue on Advanced Location-Based Services)*, vol. 31, no. 6, pp. 1114–1125, April 2008.
[8] ——, "Cognitive positioning systems," *IEEE Trans. Wireless Commun.*, vol. 6, no. 12, pp. 4475–4483, Dec. 2007.
[9] ——, "Ranging accuracy in dynamic spectrum access networks," *IEEE Commun. Lett.*, vol. 11, no. 5, pp. 405–407, May 2007.
[10] R. I. C. Chiang, G. B. Rowe, and K. W. Sowerby, "A quantitative analysis of spectral occupancy measurements for cognitive radio," in *Proc. IEEE Vehicular Technology Conf.*, Dublin, Ireland, Apr. 2007.
[11] S. Gezici, H. Celebi, H. V. Poor, and H. Arslan, "Fundamental limits on time delay estimation in dispersed spectrum cognitive radio systems," *IEEE Trans. on Wireless Commun.*, under revision, June 2008.
[12] T. A. Weiss and F. K. Jondral, "Spectrum pooling: An innovative strategy for the enhancement of spectrum efficiency," *IEEE Commun. Mag.*, vol. 42, no. 3, pp. 8–14, March 2004.
[13] S. Brandes, I. Cosovic, and M. Schnell, "Reduction of out-of-band radiation in OFDM based overlay systems," *IEEE Commun. Lett.*, vol. 10, no. 6, pp. 420–422, June 2006.
[14] H. Arslan and M. E. Sahin, "Cognitive UWB-OFDM: Pushing ultrawideband beyond its limit via opportunistic spectrum usage," *Journal of Communications and Networks*, vol. 8, no. 2, pp. 1–7, June 2006.
[15] H. V. Poor, *An Introduction to Signal Detection and Estimation*. New York: Springer-Verlag, 1994.
[16] J. G. Proakis, *Digital Communication*, 4th ed. New York: McGraw-Hill, 2001.

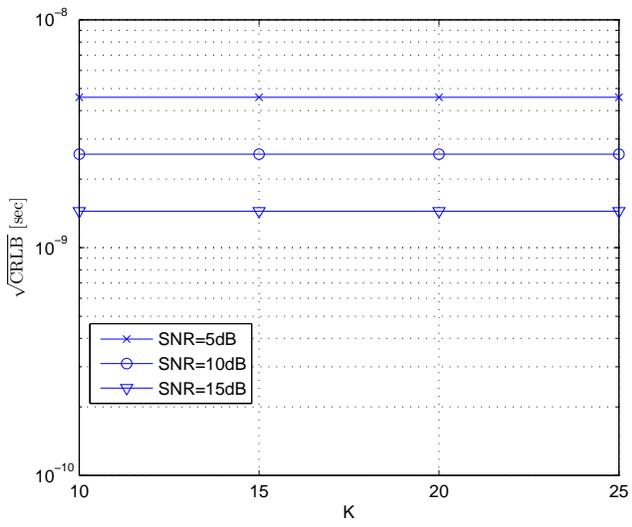

Fig. 4
$\sqrt{\text{CRLB}}$ versus $K$ for $N = 16$ and 16PSK modulation when the SNR is defined as the sum of the SNRs at different branches.

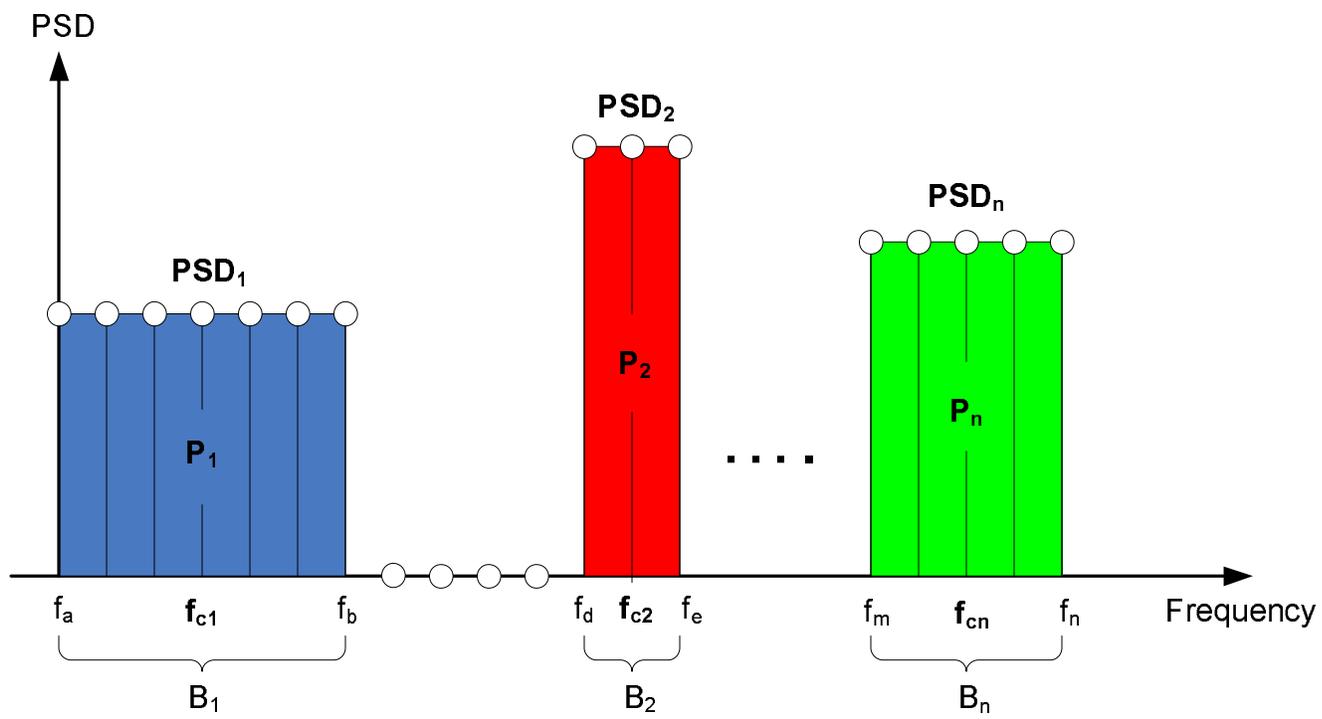

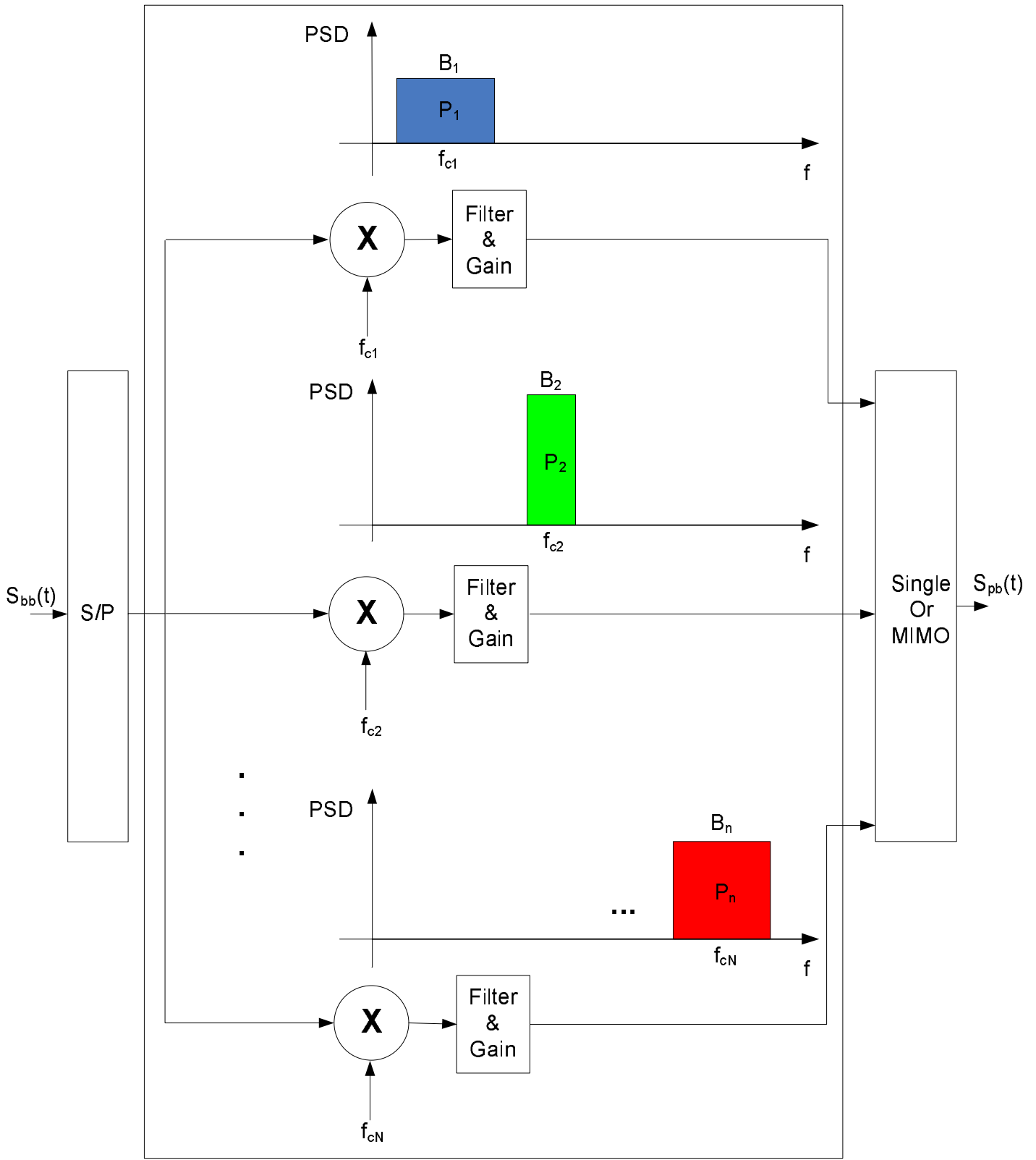

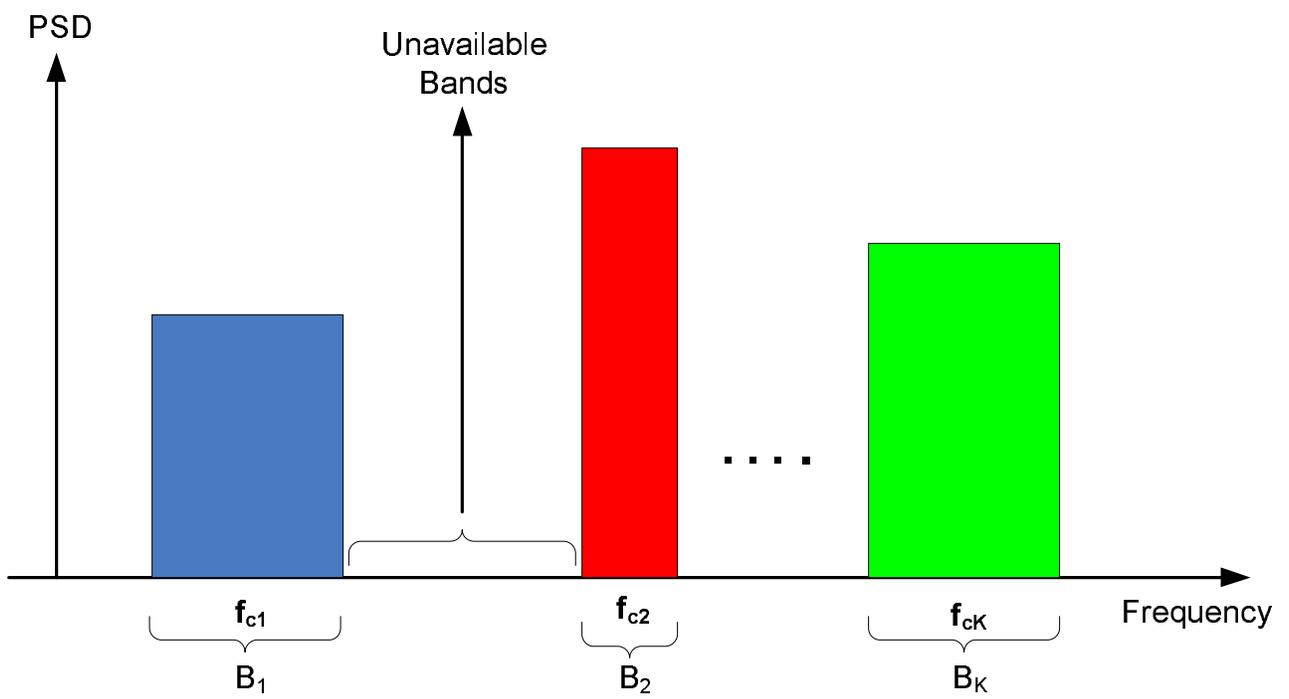